\documentclass[a4,epsfig,manuscript,aps]{revtex4}
\usepackage{graphicx}
\usepackage{multirow}

\begin{document}
\title{Rare-earth impurities in Co$_2$MnSi: \\
an opportunity to improve Half-Metallicity at finite temperatures}
\author{E. Burzo}
\affiliation{Babe\c s-Bolyai University Cluj-Napoca, RO-800084 Cluj-Napoca, Romania}
\author{L. Chioncel}
\affiliation{Institute of Theoretical 
and Computational Physics, Graz University of Technology,
A-8010 Graz, Austria}
\author{I. Balazs}
\affiliation{Babe\c s-Bolyai University Cluj-Napoca, RO-800084 Cluj-Napoca, Romania}
\author{F. Beiuseanu}
\affiliation{Faculty of Science, University of Oradea, RO-410087 Oradea, Romania}
\author{E. Arrigoni}
\affiliation{Institute of Theoretical 
and Computational Physics, Graz University of Technology,
A-8010 Graz, Austria}

\begin{abstract}
We analyse the effects of doping Holmium impurities into the
full-Heusler ferromagnetic alloy Co$_2$MnSi. Experimental results, 
as well as theoretical calculations within Density Functional Theory 
in the Local Density Approximation generalized to include Coulomb correlation
at the mean field level show 
that the holmium moment is aligned antiparallely  to that of the transition 
metal atoms. According to the electronic structure calculations,
substituting Ho on Co sites introduces a finite density of states in
the minority spin gap, while substitution on the Mn sites preserves 
the half-metallic character.
\end{abstract}

\maketitle

\section{Introduction}
An important class of materials which are intensively studied 
in the field of spintronics are the so-called half-metallic ferromagnets. 
These materials are characterized by a metallic electronic structure for 
one spin channel, whereas, for the opposite spin direction, the Fermi level 
is situated within an energy gap \cite{gr.mu.83,ka.ir.08}. From a magnetic 
point of view, these compounds can be either ferro- or ferrimagnets. 
Quite generally, theoretical calculations based on the Local-Density
Approximation (LDA) predict a perfect spin polarization at the Fermi level.   
Antiferromagentic half-metals also exits, although  this case is special since 
no symmetry relation connects spin-up and spin-down directions. The existence of 
half-metallicity in all three classes of magnetic orderings was predicted by 
{\it R. A. de Groot} and coworkers some years ago~\cite{gr.mu.83,gr.kr.86,groo.91}. 
Since then, several possible half-metallic ferromagnetic (HMF) materials have been 
theoretically investigated by ab initio calculations, and 
some of them have been synthesized and experimentally studied. 

Devices that exploit spin and charge of the electrons should operate
at not too low temperatures. Unfortunately many half-metallic systems exhibit 
a dramatic suppression of spin-polarization well below  room temperature
\cite{do.sk.03,ka.ir.08}. It is therefore important to study and to optimize 
temperature effects in  half-metallic materials. There are several effects 
potentially responsible for the suppression of spin polarisation at finite 
temperature that can be taken into account in a theoretical calculation.
The competition between hopping and local electron-electron interactions 
may produce spin disorder, which can be approximated as random intraatomic 
exchange interactions \cite{ma.an.80,li.ka.87}. These interactions modify the local 
magnetic moments and influence the spin polarization~\cite{or.fu.99,or.fu.00}.  
The exchange interactions described by a Heisenberg-type Hamiltonian
are also frequently used to discuss finite temperature magnetic effects. 
The sign and magnitude of exchange constants give information whether the 
spin structure is collinear or not \cite{li.ka.87,sa.sa.05,sa.sa.05b}. Note that 
non-collinearity may be produced also by  spin-orbit coupling. All the above 
mentioned mechanisms may be called spin-mixing effects \cite{do.sk.03,sk.do.02} 
and they may significantly influence the half-metallic character. 

It is important to point out that in all metallic ferromagnets, the 
interaction between conduction electrons and spin fluctuations determines 
their physical properties. In half-metallic ferromagnets the presence of 
the gap in one of the spin channel determines the absence of spin-flip, or 
one-magnon scattering processes which modifies considerably the electrons 
energy spectrum \cite{ed.he.73,ir.ka.83,ka.ir.08}. These electron-magnon interactions 
occur both in usual ferromagnets and in half-metals, with the difference that 
for the usual ferromagnets the states around the Fermi level are quasiparticles 
for both spin directions, while in half-metals an important role is played by 
incoherent - non-quasiparticle (NQP) - states \cite{ed.he.73,ir.ka.83}. 
%
The origin of these NQP
states is connected  with  spin-polaron processes: the
spin-down low-energy electron excitations, which are forbidden for
HMF in the one-particle picture, turn out to be possible as result of
superpositions of spin-up electron excitations and virtual magnons
\cite{ed.he.73,ir.ka.90,ir.ka.94,ka.ir.08}. 
It was demonstrated that these
states are important for spin-polarized electron spectroscopy
\cite{ir.ka.06}, nuclear magnetic resonance \cite{ir.ka.05}, 
and subgap transport in ferromagnet-superconductor junctions (Andreev reflection) 
\cite{tk.mc.01}.
Recently, the density of NQP states has been calculated using an  LDA+dynamical mean field theory  (DMFT)
approach \cite{ko.sa.06,ka.ir.08}, for several materials such as the semi-Heuslers
NiMnSb \cite{ch.ka.03} and (NiFe)MnSb \cite{ch.ar.06}, the full Heuslers Mn$_2$VAl \cite{ch.ar.09}
and Co$_2$MnSi \cite{ch.sa.08}, the zinc-blende CrAs \cite{ch.ka.05} and VAs \cite{ch.ma.06}
and the transition metal oxide CrO$_2$ \cite{ch.al.07}, demonstrating their importance 
as the essential mechanism of depolarization in half-metals.

%
%

Among the Heusler half-metals, the full-Heusler 
Co$_2$MnSi~\cite{fe.ka.06,wu.fe.06,ka.fe.06} 
has been intensively studied experimentally as well as theoretically.
This material is interesting as it shows a high Curie temperature and 
very little crystallographic disorder \cite{ra.ra.01}.
We showed  that  Co$_2$MnSi experiences a drastic spin
depolarization~\cite{ch.sa.08} with increasing temperature.
According to our calculations, depolarization is caused by the 
occurrence of non-quasiparticle states situated just above the Fermi level
in the minority spin channel \cite{ch.sa.08,ka.ir.08}. 
In order to significantly reduce depolarization
a strategy to eliminate the NQP states from the Fermi level is required. 
In other words, one should try to suppress finite-temperature magnonic 
excitations while at the same time preserving the minority spin gap.
It was suggested that the substitution of a rare-earth (R) element 
within the Mn sublattice of NiMnSb would introduce gaps in the magnonic spectrum
due to the impurity R$(4f)$ - Mn$(3d)$ coupling, possibly preserving the 
half-metallic gap~\cite{ka.ir.08,at.fa.04,chio.diss}.
In pure NiMnSb the
ferromagnetic Curie temperature $T_{C}=740$K is determined by the strength of the Mn-Mn
($3d-3d$) coupling \cite{sa.sa.05,ka.ir.08}. The strength of ($3d-4f$) couplings
in  NiMn$_{1-x}$R$_x$Sb compounds, with R=Nd, Pm, Ho, U was 
discussed previously \cite{ka.ir.08,chio.diss}. A large $3d-4f$ coupling
was obtained for the case of Nd substitution, and the weakest 
coupling was realized in the case of Ho. For all rare-earth 
substitutions the electronic-structure calculations
showed that the  half-metallic gap is not affected, and  for temperatures smaller
than the values of the $3d-4f$ couplings, interaction will lock the Mn$(3d)$ magnetic
moment fluctuation.  As a consequence, it was concluded that Nd substitution is of practical
importance for high-temperature applications \cite{ka.ir.08}, while half-metallicity with Ho 
substitutions would persist up to temperatures of around 4K as predicted by  
the calculation \cite{chio.diss,at.fa.04}.

For the case of Co$_2$MnSi, preliminary results showed that Ho can enter 
in the Co$_2$MnSi lattice~\cite{te.ch.08}.
Recently, {\it Rajanikanth et. al.} \cite{ra.ta.09} showed that substituting 
Co with $1.25 at \%$ Nd results in a single L2$_1$ phase alloy with a thin 
layer of Nd-enriched phase boundaries. A suppression of electron-magnon 
scattering was experimentally observed~\cite{ra.ta.09}.

As an ongoing investigation of impurity effects on the depolarization of 
conduction electrons in Co$_2$MnSi based alloy we analyze in this paper 
the possibility of substituting holmium in both Co or Mn lattice sites, 
as well as changes introduced in band structure of the parent compound. The 
first task is to analyze whether holmium can really replace Mn or Co
in a {\it Fm}$\bar{3}${\it m} type structure. Consequently, we prepared 
samples in which the substitution at both transition metal sites was considered.
X-rays and electron microscopy studies were carried out. Finally, the
magnetic properties and band structures of the doped compounds were analyzed.

\section{Physical properties of Ho-doped Co$_2$MnSi samples}
The Co$_2$Mn$_{1-x}$Ho$_x$Si and Co$_{2-x}$Ho$_x$MnSi alloys with nominal 
compositions $x=0.05$ and $x=0.1$ were prepared by arc melting the constituent
elements under purified argon atmosphere. These were remelted several times 
in order to ensure a good homogeneity. The alloys were annealed at $900^o C$ 
for 12 days. The X-ray analyses were performed by using a Brucker V8 diffractometer. 
Rietveld refinements were made by using the Topas code. Electron microscope 
studies were performed by using a JEOL type equipment. The compositions of the 
samples and distributions of the constituent elements were analysed. Magnetic 
measurements were made with an Oxford Instruments type equipment in fields up 
to 5 T and in the temperature range 4.2-1000 K.

\subsection{Experimental results}
\begin{figure}[h]
\includegraphics[width=0.90\linewidth]{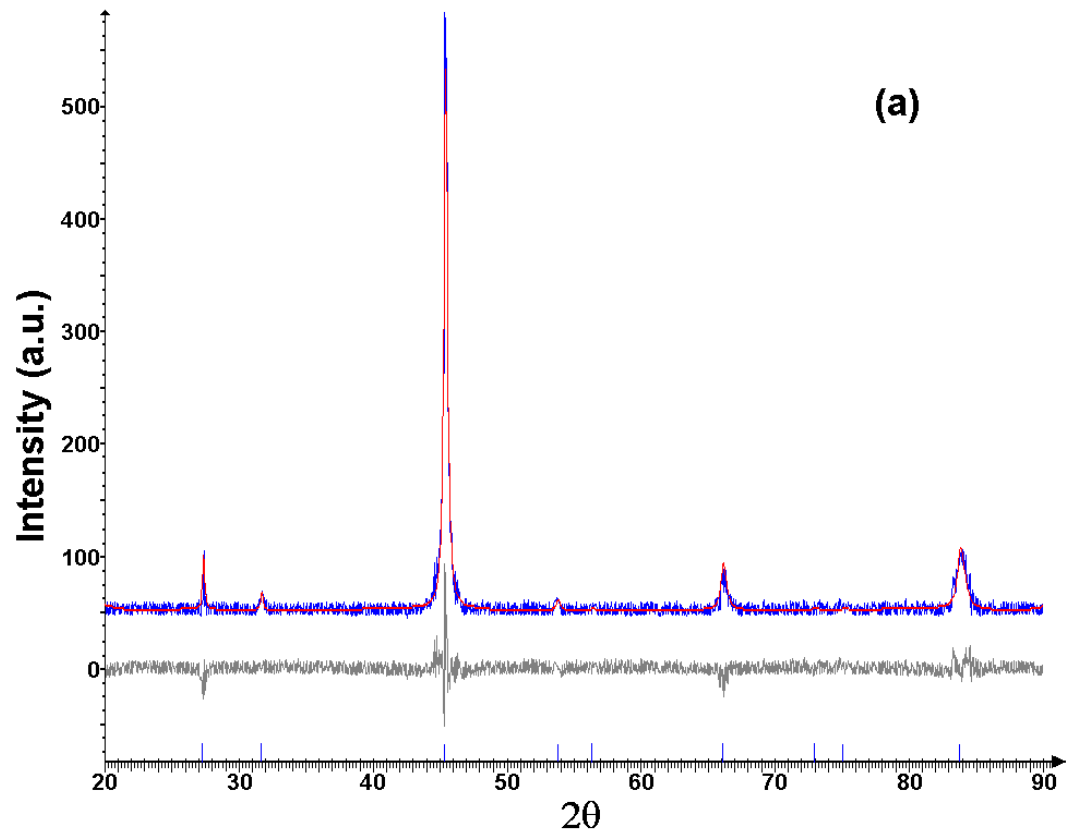}
\includegraphics[width=0.90\linewidth]{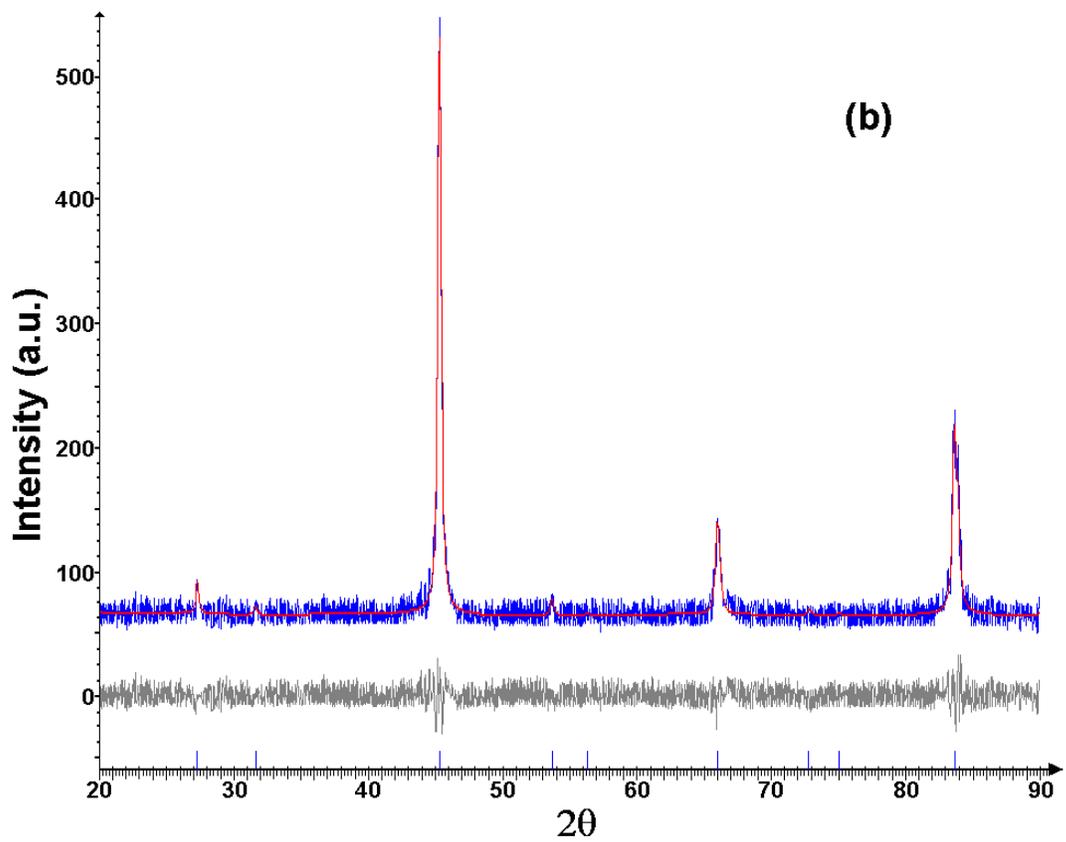}
\caption{(Color online) Refined X-ray spectra for Co$_2$Mn$_{1-x}$Ho$_x$Si 
samples with x=0.05 (a) and 0.1 (b).}\label{rx_Co2MnSi}
\end{figure}

The refined X-ray diffraction (XRD) spectra of Co$_2$Mn$_{1-x}$Ho$_x$Si 
samples with $x=0.05$  and $x=0.1$ are plotted in Fig.\ref{rx_Co2MnSi}. 
The same type of spectra were obtained when Ho substitutes Co. The alloys 
crystallize in a cubic structure having $Fm\bar{3}m$ space group. As seen 
from Table \ref{tab1}, the lattice parameters increase little when increasing 
holmium content. The linewidths of XRD spectra are a little larger in samples 
having a composition $x=0.1$ as compared to those having $x = 0.05$. This can 
be correlated with a small deformation of the lattices as the holmium content 
increases.

The better agreement between the computed spectra and those experimentally 
determined, in Co$_2$Mn$_{1-x}$Ho$_x$Si alloys is obtained when holmium
substitutes manganese located in  4c sites, according to the starting composition.
The goodness of fit (GOF) values are indeed worse (larger) if one assumes
that Ho occupies cobalt 4a or 4b sites. On the other hand, the refinements 
of the crystal structures of Co$_{2-x}$Ho$_x$MnSi samples showed GOF values 
rather close when Ho was considered either to replace cobalt or manganese.  
In addition, a further analysis of our data strongly suggests that holmium 
occupies lattice sites in Co$_2$MnSi sample, and does not sit in
interstitial positions.

\begin{figure}[h]
\includegraphics[width=0.90\linewidth]{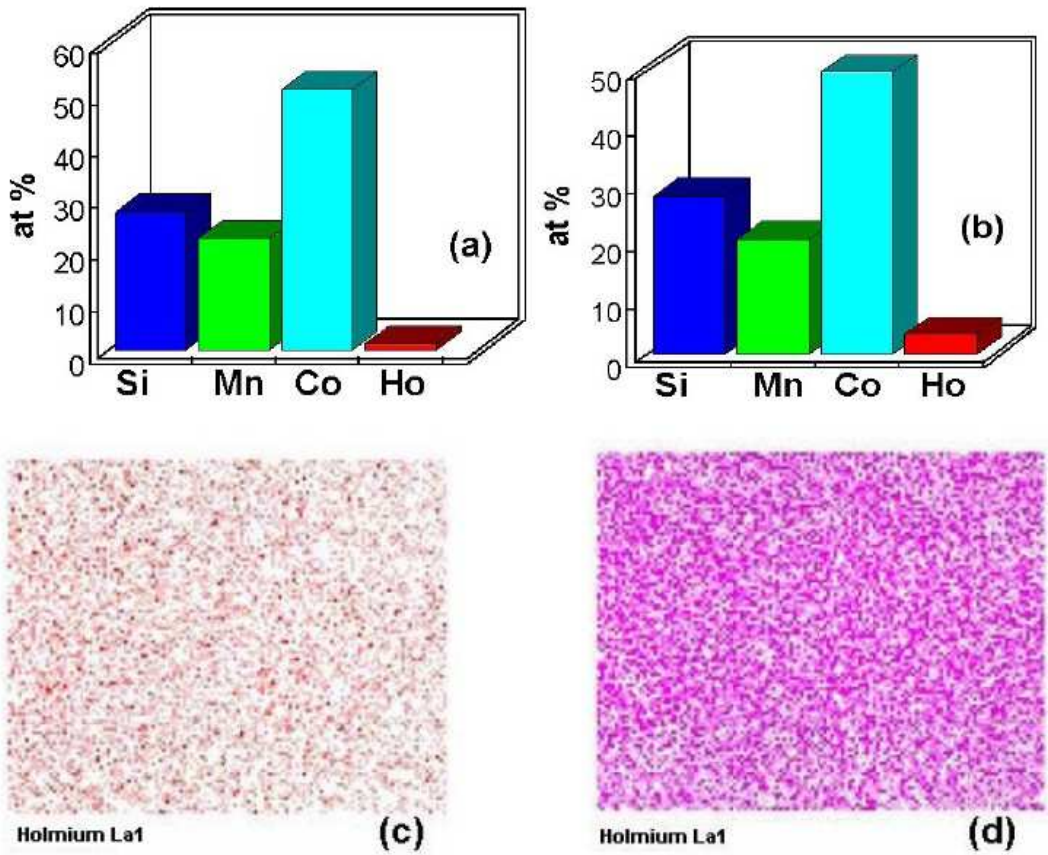}
\caption{(Color online) Compositions of the Co$_2$Mn$_{0.95}$Ho$_{0.05}$Si (a) 
and Co$_2$Mn$_{0.9}$Ho$_{0.1}$Si (b) samples as well as the holmium distributions 
in Co$_2$Mn$_{0.95}$Ho$_{0.05}$Si (c) and Co$_2$Mn$_{0.9}$Ho$_{0.1}$Si (d) samples. }\label{micro_co2mnsi}
\end{figure}

The compositions of the samples were determined by Scanning Electron 
Microscopy (SEM), for a surface area of $\approx 1 mm^2$ and a depth 
from surface up to about $200 \AA$. The determined holmium content in 
samples having initial compositions $x=0.05$ and $0.1$ were $1.30 \pm 0.15 at \%$ 
and $(2.7-2.8) \pm 0.30 at \%$ respectively in both Co$_2$Mn$_{1-x}$Ho$_x$Si 
and Co$_{2-x}$Ho$_x$MnSi systems Figs.\ref{micro_co2mnsi}a,b. These values  
are in agreement with the expected content of $1.25 at \%$ and $2.5 at \%$, 
respectively. The manganese contents in Co$_2$Mn$_{1-x}$Ho$_x$Si were
$22.8 \pm 0.7 at \% (x=0.05)$ and $22.0 \pm 0.6 at \% (x=0.1)$, while the 
cobalt content was close to $50 at \%$ - Figs.\ref{micro_co2mnsi}a,b. 
The above data suggest that holmium is distributed mainly in manganese 
sites in this compound.

On the other hand, in the case of Co$_{2-x}$Ho$_x$MnSi samples there is a decrease 
of both cobalt and manganese content as compared to expected compositions, suggesting 
that holmium can occupy both these sites. The distribution of holmium atoms in the 
samples was also analysed. As example, we show in Figs.\ref{micro_co2mnsi}c,d the 
SEM patterns of two manganese substituted samples. A random distribution 
can be seen, with no clustering effects. 

\begin{figure}[h]
\includegraphics[width=0.9\linewidth]{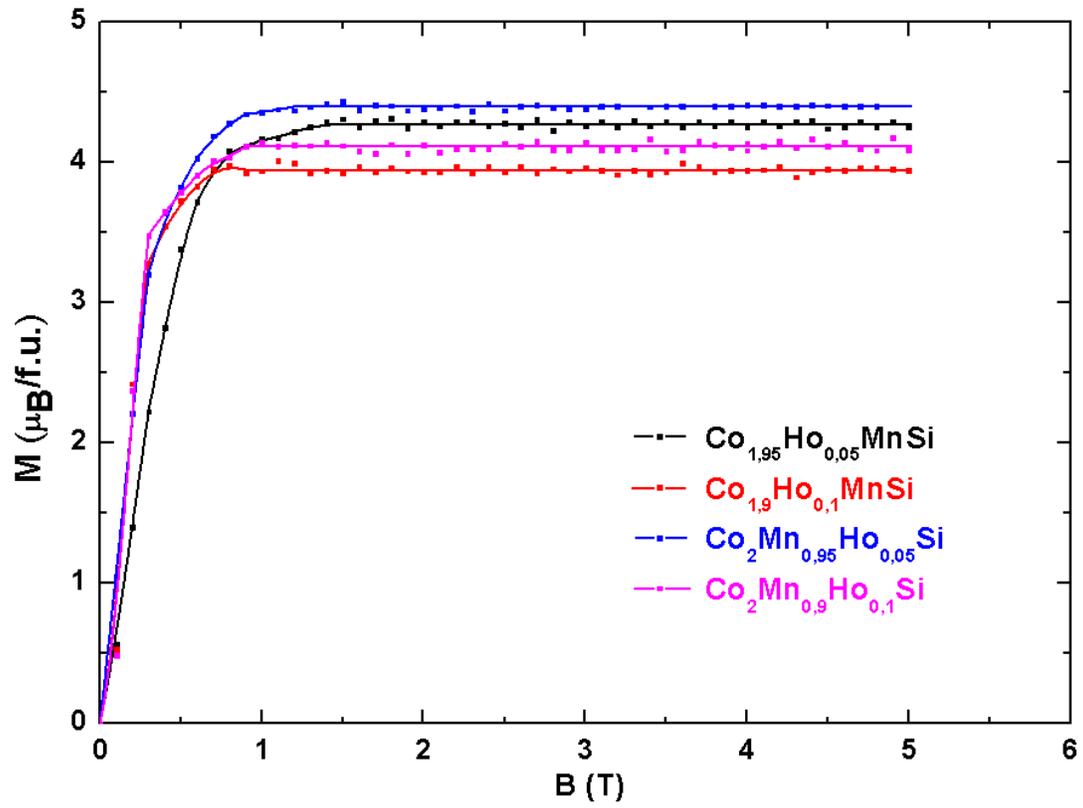}
\caption{(Color online) Magnetization isotherms at 4.2 K.}\label{mag_Co2MnSi}
\end{figure}

The magnetizations isotherms, at $4.2 K$, are shown in Fig.~\ref{mag_Co2MnSi}. The determined 
saturation magnetizations, $M$, per formula unit are given in Table~\ref{tab2}. The $M$ 
values are smaller than the ones measured in Co$_2$MnSi, namely $5\mu_B$ \cite{br.ne.00}, 
and decrease when increasing holmium content. This suggests that the holmium moments 
are antiparallely oriented to the cobalt and manganese ones. Therefore,
the magnetization measurements are in agreement with the result of structural and 
compositional analysis and confirm that holmium atoms occupy lattice sites, and that their 
moment is antiparallely oriented to cobalt and manganese ones. A better agreement 
is obtained for  a composition $x=0.05$. For samples having higher holmium content 
a little deviation from the expected stoichiometry or a possible location of Ho in 
both Co and Mn lattice sites cannot be excluded. The analysis of thermal variations 
of magnetizations in low field, shows the presence of only one magnetic phase. 
The Curie temperatures of the Ho-doped samples are only minimally reduced 
with respect to that of Co$_2$MnSi ($T_c$ = 985 K) as seen in Table~\ref{tab1}.

\subsection{Electronic Structure of Ho doped Co$_2$MnSi}
To compute the electronic structure we used the standard representation of 
the L2$_1$ structure with four inter-penetrating fcc sub-lattices. The atomic 
positions are $(0,0,0)$ for Co1, $(1/2,1/2,1/2)$ for  Co2, and
$(1/4,1/4,1/4)$ and $(3/4,3/4,3/4)$ for  Mn and Si, respectively. 
We note that the Co1 and Co2 sites are equivalent. We used the experimental 
lattice constant $a=5.65\AA$ and the linear muffin-tin orbital method (LMTO)\cite{an.je.84}
extended to include the mean-field Hubbard +U approximation (LDA+U) \cite{an.ar.97l}.
It is generally known that the LDA+U method applied to metallic compounds containing rare-earth 
elements, gives a qualitative improvement compared with the LDA not only for excited-state 
properties such as energy gaps but also for ground-state properties such as magnetic 
moments and interatomic exchange parameters. Therefore it is expected that it will 
provide a correct description for the narrow $Ho(4f)$ states, the minority spin gap 
and the $3d-4f$ coupling in the Ho doped Co$_2$MnSi.

\begin{figure}[h]
\includegraphics[width=0.9\linewidth]{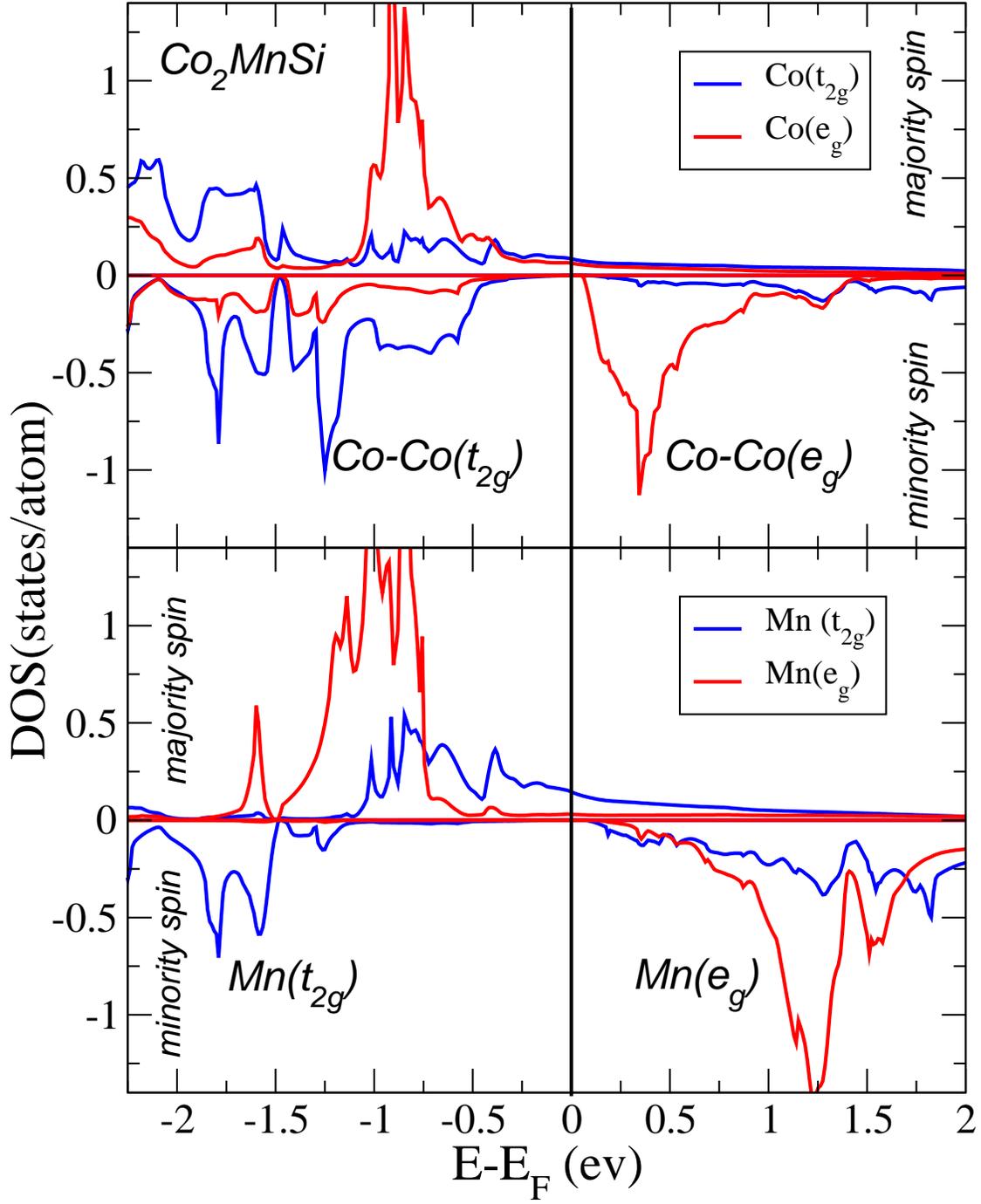}
\caption{(Color online) Spin and orbital resolved density of states of the
(stoichiometric) Co$_2$MnSi full-Heusler alloy obtained by spin-polarized LDA. The   
half-metallic gap in the minority spin channel is formed between the hybrid 
Co-Co(t$_{2g}$) and Co-Co (e$_{g}$) orbitals. No significant Mn-orbitals 
contributions to the edges of the minority spin gap can be seen in the lower 
panel.}
\label{gap_Co2MnSi}
\end{figure}

The origin of the minority band gap in the full Heusler alloy was discussed by 
{\it Galanakis et al.}~\cite{ga.ma.06}. Based on the LDA band-structure calculations 
for the undoped Co$_2$MnSi, it was shown that Co1 and Co2 couple forming 
bonding/anti-bonding hybrids. The Co1(t$_{2g}$/e$_{g}$) - Co2(t$_{2g}$/e$_g$)  
hybrid bonding orbitals hybridize with the Mn(t$_{2g}$/e$_{g}$) manifold,
while the Co-Co hybrid anti-bonding orbitals remain uncoupled due to their symmetry.
The Fermi energy is situated within the minority spin gap formed by the triply 
degenerate anti-bonding Co-Co(t$_{2g}$) and the double degenerate anti-bonding 
Co-Co(e$_g$) (Fig. \ref{gap_Co2MnSi}). Due to the orbitals occupation, the majority 
spin channel has a metallic character, with a small density of states at the Fermi 
level. In agreement with previous~\cite{ga.ma.06,sa.sa.05,sa.sa.05b,ku.fe.07} calculations, 
an integer total spin magnetic moment of $5.00\mu_B$, and a minority spin gap of 
magnitude $0.42eV$ is obtained. 

The behavior of the magnetic moment as a function of temperature  in undoped Co$_2$MnSi
was discussed in the framework of the LDA+DMFT method~\cite{ch.sa.08}. Both magnetisation 
and spin polarisation decrease with increasing temperature, although the latter decreases
much faster. Above the Fermi level, non-quasiparticle (NQP)
states~\cite{ed.he.73,ir.ka.90,ir.ka.94,ka.ir.08} are present, which influence significantly 
the finite temperature properties~\cite{ch.sa.08}. 

\begin{figure}[h]
\includegraphics[width=0.9\linewidth]{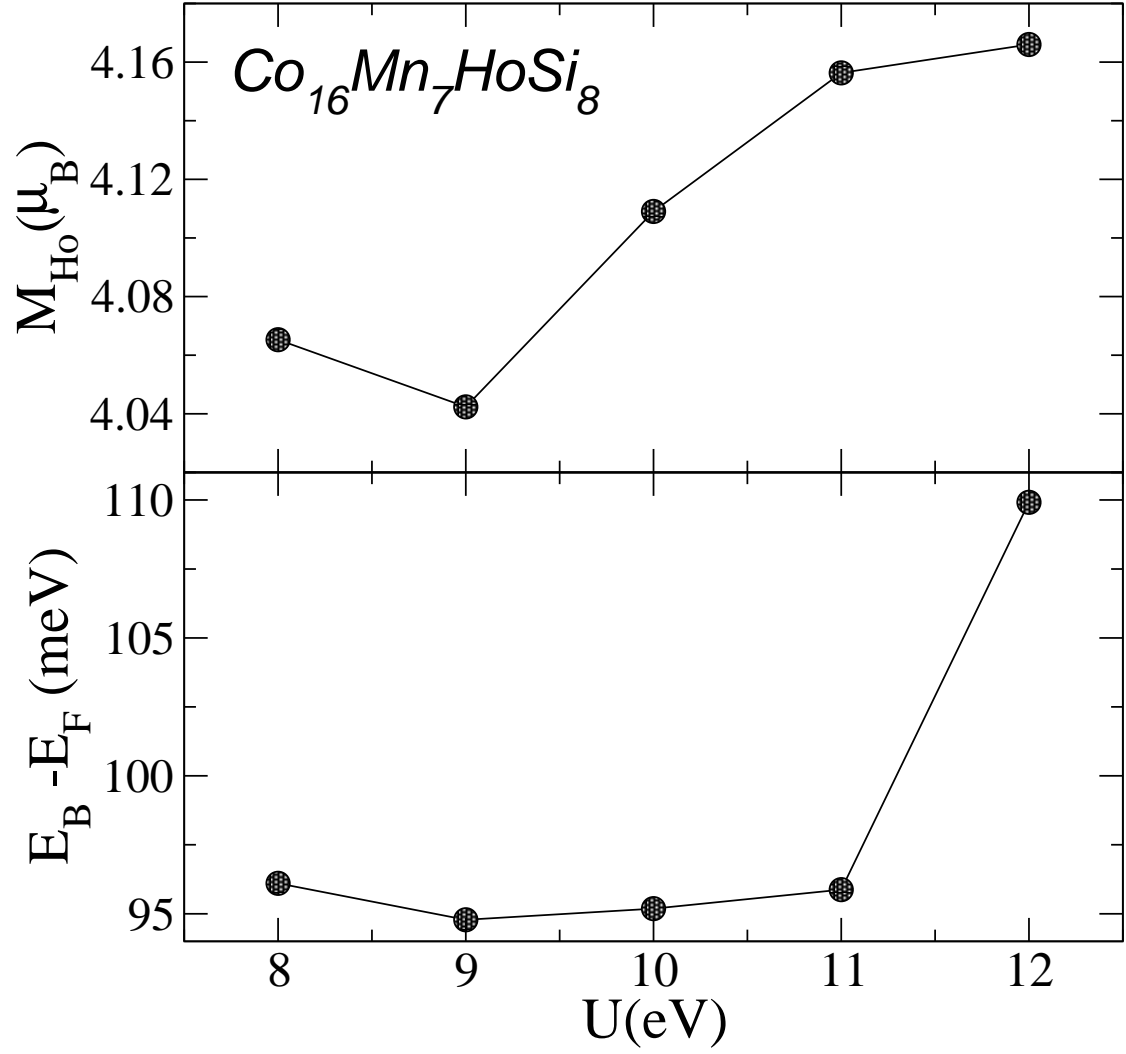}
\caption{(Color online) The minority spin 
half-metallic gap as function of the average Coulomb  parameter U.
No significant U dependence is present which demonstrates the lack on 4f-3d 
hybridizations around the Fermi level.}
\label{gap_U}
\end{figure}

\begin{figure}[h]
\includegraphics[width=0.9\linewidth]{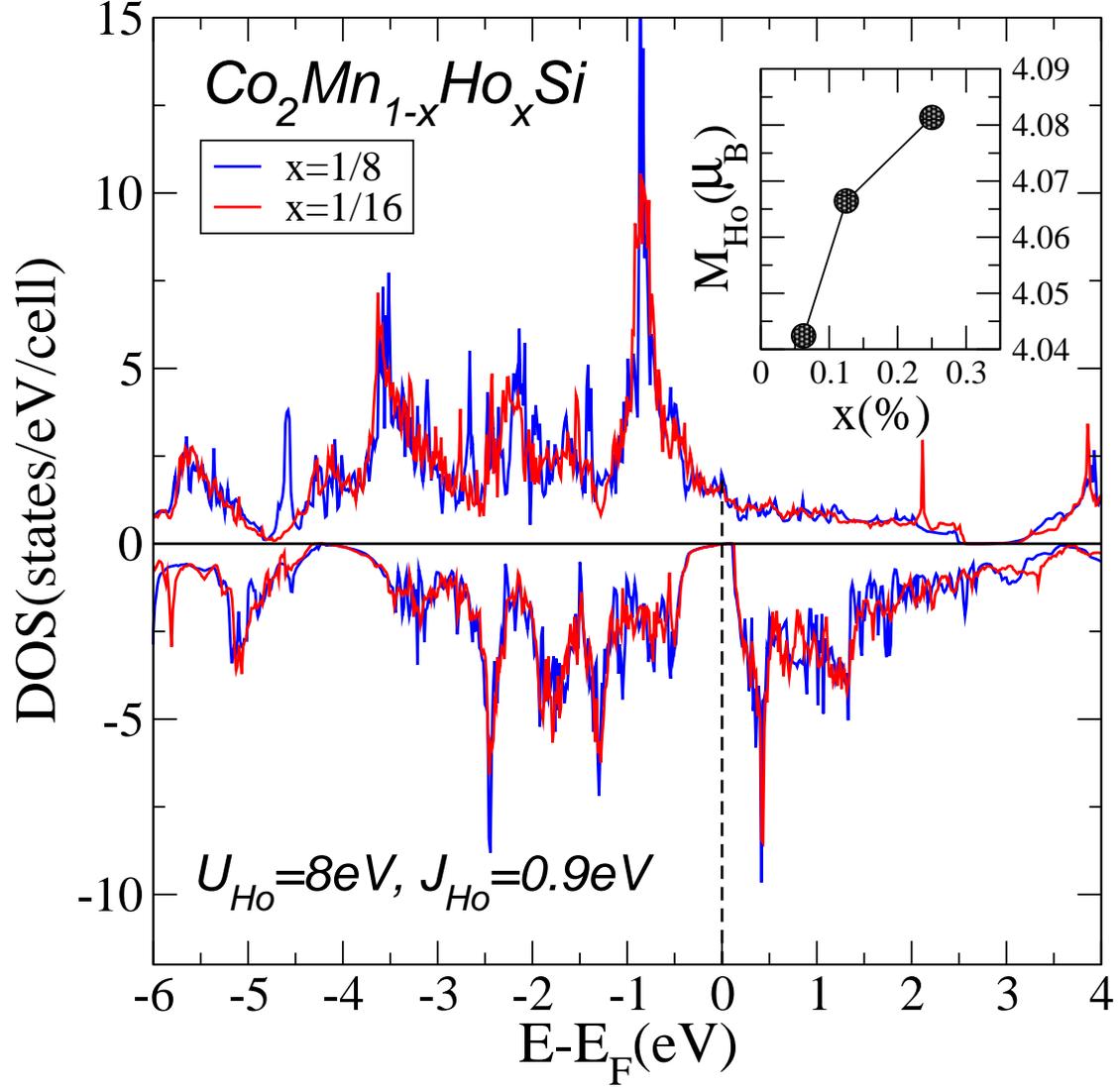}
\caption{(Color online) Spin resolved density of states for different
eight respectively 16 times larger supercells of Co2MnSi. The inset shows the 
composition dependence Ho magnetic spin moments. No essential changes are 
visible around the Fermi level.}
\label{gap_conc}
\end{figure}

We have performed LDA+U calculations for
different values of the average Coulomb interaction parameter U and different
sizes of the unit cell, in order to investigate the sensitivity of the results
with respect to the these parameters. The parameters study was performed for the
energetically most favorable Ho substitution, namely the replacement of a Mn
atom by Ho with an antiparallel coupling of the Ho spin moment with respect to the
Mn itinerant electrons spin moment. For all our values of U a half-metallic
solution is obtained with a minority gap having similar width as in the spin
polarized LDA calculation, however depending on the strength of U the Fermi level
moves towards the middle of the gap.
In Fig. \ref{gap_U} we present the computed value of the Ho spin moments together
with the difference between the bottom of the conduction band ($E_B$) and the 
Fermi level $E_B-E_F$ as a function of the average coulomb parameter $U$. As 
one can see there is no significant dependence in the studied range of the U 
parameter $8-12eV$: a slight increase of the magnetic moment from $4.04$ up 
to $4.15\mu_B$ is visible, while the maximum $E_B-E_F$ difference is of order 
of $100meV$. Therefore in the following calculations we take the values U$_{Ho}=8eV$ and 
$J_{Ho}=0.9eV$ for the Coulomb and exchange parameters, which agrees with the 
values reported in literature for metallic rare-earth compounds \cite{an.ar.97l}. 
Using these values we performed calculations
for three supercells containing 16 atoms in  the case of
Co$_8$HoMn$_3$Si$_4$, 32 atoms for Co$_{16}$Mn$_7$HoSi$_8$ and respectively
64 atoms for the Co$_{32}$Mn$_{15}$HoSi$_{16}$  cell. In Fig. \ref{gap_conc}
we shows the spin resolved density of states for the two largest supercells.
In the inset the concentration dependence of the Ho spin moment is presented
for all three supercells. 
Again one can see that the results does no significantly change with the dimension 
of the cells. Therefore for computational convenience,  
and in order to obtain the value for the orbital magnetic moment of Ho, we
carried out the LDA+U calculation including both spin-orbit coupling 
and non-collinearity effects for the smallest supercell in discussion Co$_8$HoMn$_3$Si$_4$. 
Our calculation yields a self-consistent collinear antiparallel
configuration of the holmium moments with respect to the transition 
metal moments, with values of spin and orbital moments of $3.917 \mu_B$ 
and $5.914 \mu_B$, respectively. The total magnetic moment ($9.83 \mu_B$) 
is close to the usual value determined by neutron diffraction in compounds having 
cubic structure with similar space group, such as HoFe$_2$~\cite{bu.ch.90}. 
Similarly to the Ho spin moments, which were shown not to change significantly 
form one supercell to the other,   
we do not expect significant changes of the Ho orbital moments as well,
since the magnitude of Ho spin-orbit coupling does not change appreciably 
with concentration, provided  one uses the same type of structure and 
similar lattice parameters. For this reason, it is justified to neglect 
spin-orbit coupling for the band structure calculation in the larger supercell 
presented below. In this case, in order to compare the total magnetic moment 
with the experimentally measured values, the orbital magnetic moment calculated 
above for the smaller supercell is then added to the spin moments to obtain the 
results shown in Table~\ref{tab2}.

\begin{figure}[h]
\includegraphics[width=0.80\linewidth]{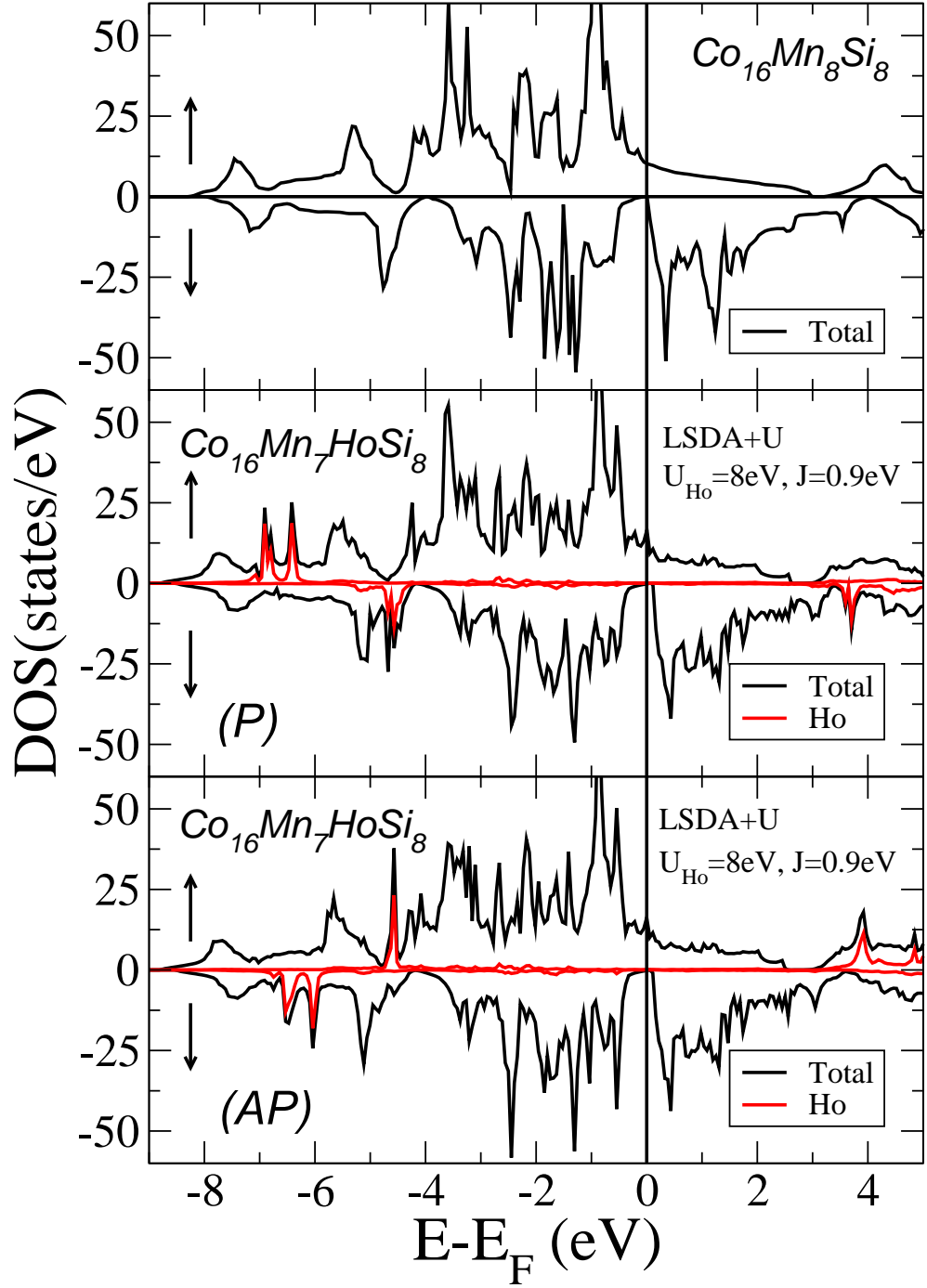}
\caption{(Color online) Spin resolved density of states of the Co$_{16}$Mn$_{7}$HoSi$_8$
supercell in the  parallel (P) and anti-parallel (AP) configuration between
$Ho(4f)-Mn(3d)$ moments. For comparison, the Co$_2$MnSi density of states is
also plotted. In the presence of Ho, Fermi level shifts towards the middle of the 
minority-spin gap, however the magnitude of the gap remains unchanged.}
\label{dos-HoMn_Co2MnSi}
\end{figure}

As discussed above, in the case of Co$_2$MnSi rare-earth impurities can in principle 
enter in both the Co and the Mn- sublattices. 
For this reason, we present results for the electronic-structure 
calculations carried out on the Co$_{16}$Mn$_7$HoSi$_8$ supercell to simulate Ho 
substitution at the Mn sites, and for a Co$_{15}$HoMn$_8$Si$_8$ cell when Ho replaces Co. 
The self-consistent calculations were performed assuming  a collinear,
parallel and respectively anti-parallel orientation of the Ho$(4f)$ magnetic 
moment with respect to the Mn/Co$(3d)$ ones. The resulting density of states 
for Co$_{16}$Mn$_7$HoSi$_8$ with both parallel and antiparallel 
orientation between Ho$(4f)$-Mn$(3d)$ moments are plotted in Fig. \ref{dos-HoMn_Co2MnSi}. 
The half-metallic state is stable in both cases, with a gap
of similar magnitude. However, the presence of rare-earth impurities induces a
redistribution of states such that the Fermi level moves slightly towards
the center of the gap, thus making the half-metallic state more stable.   
It is important to mention that in the case of Co$_{16}$Mn$_7$HoSi$_8$ 
Ho$(4f)$-orbitals do not hybridize with the $3d$ orbitals near the Fermi
level, the behaviour of DOS near $E_F$  is very similar to the 
undoped case, so the nature of carriers around $E_F$ is not changed. 

To compute the values of the Ho$(4f)$-Mn$(3d)$ exchange coupling, we use 
a two-sublattice model that describes the sublattice of Ho$(4f)$ spin moments 
as being antiparalely or parallely oriented to the Mn$(3d)$ spins sublattice. 
In this simplified model the exchange coupling, J, corresponds to the 
intersublattice couplings. In this model the Mn$(3d)$  and the Ho$(4f)$-states are   
treated within a  mean field LDA+U approach, whereas 
the Ho$(4f)$-Mn$(3d)$ interaction was treated as a perturbation. 
The corresponding mean-field Hamiltonian can be written in the form:
\begin{equation} 
H \approx H_{LDA+U} -J \sum_{i,\delta} \sigma^{3d}_i S^f_{i+\delta}
\end{equation}
Here, $\sigma^{3d}_i$ represents the spin operator of the 3d electrons at site ${\bf r_i}$,
and $S^f_{i+\delta}$  the spin of the 4f shell at the site ${\bf r_{i+\delta}}$.
Within this model the Mn$(3d)$ local moment fluctuation could be quenched by a 
strong Ho$(4f)$-Mn$(3d)$  coupling affecting the magnon excitations  
of the $3d$ conduction electron spins of the Mn-sublattice. 
Given the geometry of the cell, when Ho is  substituted into the Mn- sublattice, 
twelve pairs of Ho$(4f)$-Mn$(3d)$ are formed. The corresponding Ho$(4f)$-Mn$(3d)$
coupling constant is calculated as the energy difference between the
parallel $E_{P}$ and the antiparallel $E_{AP}$ configuration. In this way, we 
obtain a value of $J \simeq 88 K$. In analysing the behaviour of exchange constants
in Co$_2$MnSi, it was recently pointed out that Co-Mn interactions are responsible for 
the stability of the ferromagnetism \cite{sa.sa.05,sa.sa.05b}, in particular the main exchange 
parameter corresponds to the nearest-neighbour Co1-Mn interaction which already gives 
$70\%$ of the of the total contribution to J being about ten times larger than the 
Co-Co and Mn-Mn interactions \cite{ku.dr.05,ku.fe.07}. For the small Ho content in the 
Co$_2$MnSi,  it is shown experimentally , see table Tab.\ref{tab1}, that the ferromagnetic Curie 
temperature is not significantly reduced, so one can conclude that the Co1-Mn interaction 
is not considerably affected, and in particular this interaction dominates over the $3d-4f$ 
coupling.  

The resulting total magnetic moments of the 
supercell, taking into account  orbital Ho moments as well, 
were $26.09 \mu_B$ and $45.91 \mu_B$ for antiparallel and parallel 
orientations of Ho with respect to the Mn and Co moments, respectively.

As shown in Ref. \cite{ga.ma.06} and briefly described above 
(Fig. \ref{gap_Co2MnSi}), the minority gap is formed by the triply 
degenerate Co-Co anti-bonding t$_{2g}$ and the double degenerate Co-Co 
anti-bonding e$_g$. Thus, it is expected that a substitution in the Co sublattice 
would have a stronger effect on the electronic structure than the substitution 
in the Mn sublattice, described above.   

\begin{figure}[h]
\includegraphics[width=0.90\linewidth]{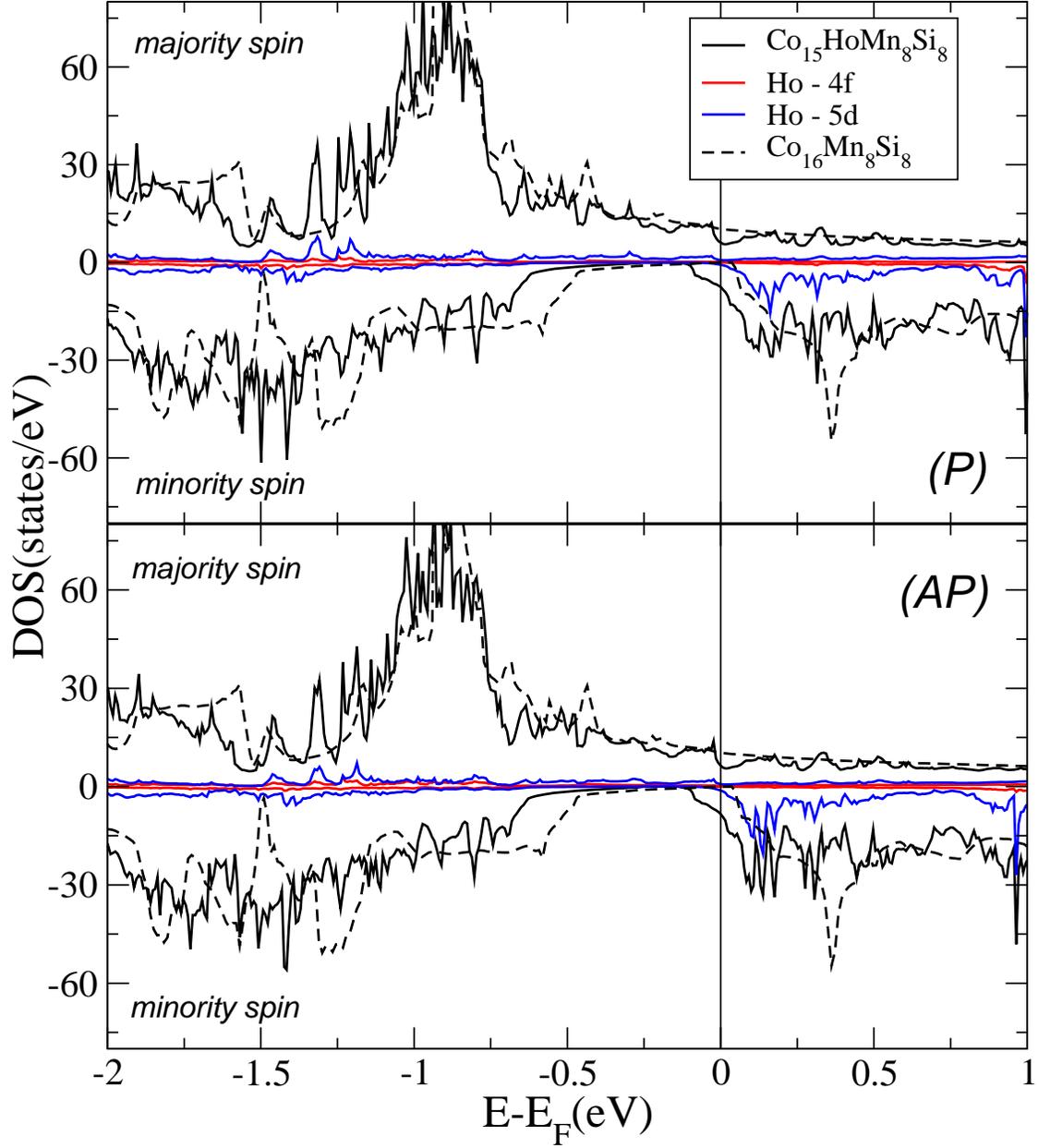}
\caption{(Color online) Spin resolved density of states around the Fermi level
for the Co$_{15}$HoMn$_8$Si$_8$ supercell, in the parallel and anti-parallel 
configurations, where Ho is substituted into  Co1 lattice sites. 
The Ho-4f and -5d states are magnified by a factor of ten in 
order to evidence hybridzation effects. The minority-spin gap is closed
due to the Co$^{3d}$(e$_g$)-Ho$^{5d}$($e_g$) anti-bounding coupling which 
pushes the Fermi level within the conduction band.}
\label{dos-HoCo1_Co2MnSi}
\end{figure}

Results of the selfconsistent spin polarized LDA+U calculation for the Co$_{15}$HoMn$_8$Si$_8$ supercell, 
in which Ho atoms substitutes one of the cobalt sites is shown in Fig.\ref{dos-HoCo1_Co2MnSi}. 
Identical results are obtained when Ho substitutes Co1 or Co2 because 
cobalt occupies equivalent sites.  
The  magnetic moments per formula unit in the doped material show
a departure from the integer values corresponding to the half-metallic case.
For Ho impurity on the cobalt site, neglecting the orbital contribution, in 
the anti-parallel $4f-3d$ configuration we obtain $\mu^{(AP)}_{Ho/Co1}=34.81\mu_B$ 
per supercell, while for the parallel case we have $\mu^{(P)}_{Ho/Co1}=41.87\mu_B$ 
per supercell. Also in this case, the antiparallel configuration is the most stable
one, in agreement with the magnetic measurements discussed above. 
An interesting aspect is that, in contrast to the case of substitution
at the Mn site, substitution at Co sites fills the minority-spin 
gap, as can be seen in Fig. \ref{dos-HoCo1_Co2MnSi}.
Notice that Ho$(4f)$ and $(5d)$ states shown in Fig. \ref{dos-HoCo1_Co2MnSi} were
multiplied by ten in order to evidence possible hybridization effects. 
As one can see (Fig. \ref{dos-HoCo1_Co2MnSi}) the Ho$(4f)$ states show no 
significant contribution in a large energy range ($E_F \pm 3eV$) around 
the Fermi level. Occupied and empty Ho$(4f)$ states are present in the energy 
range -8 to -6eV, and 2 to 4eV respectively, not shown in Fig. \ref{dos-HoCo1_Co2MnSi}. 
The anti-bonding Co$^{3d}$($e_g$)-Ho$^{5d}$($e_g$) states appear 
for both parallel and anti-parallel configurations above the Fermi level
up to energies $E_F+0.2eV$ and these states determine the shift of the 
Fermi level within the conduction band, thus the closure of the 
half-metallic gap. 

Notice that as result of hybridization effects, a small negative 
magnetic moment is induced on silicon. Values of $-0.017 \mu_B$ are 
obtained when Ho is located on Mn sites and $-0.047 \mu_B$ when Ho replaces Co. 
The computed magnetic moments of the atoms when Ho replaces Mn or Co in both 
the antiparallel and parallel alignments are listed in Table~\ref{tab2}. 
The computed magnetic moments per formula unit for 
Co$_2$Mn$_{7/8}$Ho$_{1/8}$Si and Co$_{15/8}$MnHo$_{1/8}$Si are also shown. These 
compare reasonably with experimental data, when corrections for different 
compositions are made. 
This is done by extrapolating the experimental data obtained with 
$x=0.05$ and $0.10$ to the theoretically computed case $x=0.125$.
A rather good agreement between computed and experimental values is obtained 
when the holmium replaces cobalt atoms in the antiparallel configuration, while 
a worse agreement is shown when holmium replaces manganese, see Tab.\ref{tab2},
although energetically substitution in the former way is less favorable.  

\section{Conclusion} 
The behavior of  spin polarization as a function of temperature is an important issue
for many spintronic materials. In our recent paper \cite{ch.sa.08},
we interpreted the measured drastic depolarization in Co$_2$MnSi as
due to the existence of non-quasiparticle states just above the Fermi
level. Since the origin of these states is related to the electron-magnon 
interaction, a possibility to reduce this depolarisation effect  would 
be to quench the magnon excitations, while keeping the minority spin channel gapped. 
This possibility was already discussed in connection to light-rare earth 
(Nd) insertion into the host  Co$_2$MnSi~\cite{ra.ta.09}, supporting the validity
of the proposed scenario. In the present work, we first demonstrate experimentally the 
possibility that a heavy rare-earth element (Ho) enters into the transition-metal 
sublattices in Co$_2$MnSi. In addition, we carried out an electronic-structure 
calculation whose results suggest that the half-metallicity is 
maintained when Ho replace Mn atom in an anti-parallel magnetic Ho(4f)-Mn(3d) 
configuration. Since the calculated exchange coupling $J$ turns out to be about $88 K$, we
expect that for temperatures smaller than $J$ such a substitution could in principle 
influence the magnonic excitations, while at the same time leaving the half-metallic gap
unchanged. On the other hand, our results show that when the substitution 
takes place at Co sites, the minority-spin gap is filled, and half-metallicity is lost.  

The present calculations does not take into account relaxation effects which might appear
as a result of substitution. 
This  will constitute the subject of future investigations. 
Dynamic correlation effects 
are also expected to be important. In particular, it would be
interesting to investigate up to which extent magnonic excitations and
electron-magnon interactions are affected by Ho substitution. We plan
to investigate this effect in the future.

M.I. Katsnelson and A.I. Lichtenstein are acknowledged for useful discussions.
LC and EA acknowledge financial support by the Austrian science fund (FWF
project P18505-N16). 

\bibliographystyle{prsty}
\bibliography{references_database}

\begin{thebibliography}{10}

\bibitem{gr.mu.83}
R.~A. de~Groot, F.~M. Mueller, P.~G. van Engen, and K.~H.~J. Buschow, Phys.
  Rev. Lett. {\bf 50},  2024  (1983).

\bibitem{ka.ir.08}
M.~I. Katsnelson {\it et~al.}, Reviews of Modern Physics {\bf 80},  315
  (2008).

\bibitem{gr.kr.86}
R.~A. de~Groot, A.~M. van~der Kraan, and K.~H.~J. Buschow, J. Magn. Magn.
  Mater. {\bf 61},  330  (1986).

\bibitem{groo.91}
R.~A. de~Groot, Physica B {\bf 172},  45  (1991).

\bibitem{do.sk.03}
P.~A. Dowben and R. Skomski, Journal of Applied Physics {\bf 93},  7948
  (2003).

\bibitem{ma.an.80}
A.~R. Machintosh and O.~K. Andersen,  in {\em Electrons at the Fermi surface},
  edited by M. Springford (Cambridge Univ. Press., London, 1980).

\bibitem{li.ka.87}
A.~I. Liechtenstein, M.~I. Katsnelson, V.~P. Antropov, and V.~A. Gubanov, J.
  Mag. Mag. Matt. {\bf 67},  65  (1987).

\bibitem{or.fu.99}
D. Orgassa, H. Fujiwara, T.~C. Schulthess, and W.~H. Butler, Phys. Rev. B {\bf
  60},  13237  (1999).

\bibitem{or.fu.00}
D. Orgassa, H. Fujiwara, T.~C. Schulthess, and W.~H. Butler, J. Appl. Phys.
  {\bf 87},  5870  (2000).

\bibitem{sa.sa.05}
E. Sasioglu, L.~M. Sandratskii, and P. Bruno, J. Phys.: Condens. Matter {\bf
  17},  995  (2005).

\bibitem{sa.sa.05b}
E. Sasioglu, L.~M. Sandratskii, P. Bruno, and I. Galanakis, Phys. Rev. B {\bf
  72},  184415  (2005).

\bibitem{sk.do.02}
R. Skomski and P.~A. Dowben, Europhys. Lett., {\bf 58},  544  (2002).

\bibitem{ir.ka.83}
V.~Y. Irkhin and M.~I. Katsnelson, Sov. Phys. - Solid State {\bf 25},  1947
  (1983).

\bibitem{ed.he.73}
D.~M. Edwards and J.~A. Hertz, Journal of Physics F-Metal Physics {\bf 3},
  2191  (1973).

\bibitem{ir.ka.90}
V.~Y. Irkhin and M.~I. Katsnelson, J. Phys.: Condens. Matter {\bf 2},  7151
  (1990).

\bibitem{ir.ka.94}
V.~Y. Irkhin and M.~I. Katsnelson, Phys. Usp. {\bf 37},  659  (1994).

\bibitem{ir.ka.06}
V.~Y. Irkhin and M.~I. Katsnelson, Phys. Rev. B {\bf 73},  104429  (2006).

\bibitem{ir.ka.05}
V.~Y. Irkhin and M.~I. Katsnelson, Eur. Phys. J. B {\bf 43},  479  (2005).

\bibitem{tk.mc.01}
G. Tkachov, E. McCann, and V.~I. Fal\char39{}ko, Phys. Rev. B {\bf 65},  024519
   (2001).

\bibitem{ko.sa.06}
G. Kotliar {\it et~al.}, Rev. Mod. Phys. {\bf 78},  865  (2006).

\bibitem{ch.ka.03}
L. Chioncel, M.~I. Katsnelson, R.~A. de~Groot, and A.~I. Lichtenstein, Phys.
  Rev. B {\bf 68},  144425  (2003).

\bibitem{ch.ar.06}
L. Chioncel, E. Arrigoni, M.~I. Katsnelson, and A.~I. Lichtenstein, Phys. Rev.
  Lett. {\bf 96},  137203  (2006).

\bibitem{ch.ar.09}
L. Chioncel, E. Arrigoni, M.~I. Katsnelson, and A.~I. Lichtenstein, Phys. Rev.
  B {\bf 79},  125123  (2009).

\bibitem{ch.sa.08}
L. Chioncel {\it et~al.}, Phys. Rev. Lett. {\bf 100},  086402  (2008).

\bibitem{ch.ka.05}
L. Chioncel {\it et~al.}, Phys. Rev. B {\bf 71},  085111  (2005).

\bibitem{ch.ma.06}
L. Chioncel {\it et~al.}, Phys. Rev. Lett. {\bf 96},  197203  (2006).

\bibitem{ch.al.07}
L. Chioncel {\it et~al.}, Phys. Rev. B {\bf 75},  140406  (2007).

\bibitem{fe.ka.06}
G.~H. Fecher {\it et~al.}, Appl. Phys. Lett. {\bf 99},  08J106  (2006).

\bibitem{wu.fe.06}
S. Wurmehl {\it et~al.}, Journal of Applied Physics {\bf 99},  08J103  (2006).

\bibitem{ka.fe.06}
H.~C. Kandpal, G.~H. Fecher, C. Felser, and G. Sch\"{o}nhense, Phys. Rev. B
  {\bf 73},  094422  (2006).

\bibitem{ra.ra.01}
M.~P. Raphael {\it et~al.}, Appl. Phys. Lett. {\bf 79},  4396  (2001).

\bibitem{at.fa.04}
J.~J. Attema {\it et~al.}, J. Phys.: Condens. Matter {\bf 16},  S5517  (2004).

\bibitem{chio.diss}
L. Chioncel, Finite Temperature Electronic Structure, beyond Local Density
  Approximation, PhD thesis, University Nijmegen, 2004.

\bibitem{te.ch.08}
R. Tetean {\it et~al.}, Appl. Surf. Science {\bf 255},  685  (2008).

\bibitem{ra.ta.09}
A. Rajanikanth, Y.~K. Takahashi, and K. Hono, Journal of Applied Physics {\bf
  105},  063916  (2009).

\bibitem{br.ne.00}
P.~J. Brown, K.~U. Neumann, P.~J. Webster, and K.~R.~A. Ziebeck, J. Phys.:
  Condens. Matter {\bf 12},  1827  (2000).

\bibitem{an.je.84}
O.~K. Andersen and O. Jepsen, Phys. Rev. Lett. {\bf 53},  2571  (1984).

\bibitem{an.ar.97l}
V.~I. Anisimov, F. Aryasetiawan, and A.~I. Lichtenstein, Journal of Physics:
  Condensed Matter {\bf 9},  767  (1997).

\bibitem{ga.ma.06}
I. Galanakis, P. Mavropoulos, and P.~H. Dederichs, J. Phys. D: Appl. Phys. {\bf
  39},  765  (2006).

\bibitem{ku.fe.07}
J. K\"{u}bler, G.~H. Fecher, and C. Felser, Phys. Rev. B {\bf 76},  024414
  (2007).

\bibitem{bu.ch.90}
E. Burzo, A. Chelkowski, and H.~R. Kirchmayr,  in {\em Landolt B\"ornstein
  Handbook} (Springer Verlag, Heidelberg, 1990), Vol.~19d2, p.\ 130.

\bibitem{ku.dr.05}
Y. Kurtulus, R. Dronskowski, G.~D. Samolyuk, and V.~P. Antropov, Phys. Rev. B
  {\bf 71},  014425  (2005).

\end{thebibliography}

\newpage

\begin{table}[h]
\begin{tabular}{|c|c|c|c|}
\hline
 Alloy & Lattice constant     & Saturation magnetization at 4.2K & $T_c$ \\
       &     ($\AA$)          &         $(\mu_B/fu)$            & $(K)$ \\ \hline\hline
$Co_2Mn_{0.95}Ho_{0.05}Si   $ & 5.651(2) & $4.40 \pm 0.07$  & 984  \\
$Co_2Mn_{0.90}Ho_{0.10}Si   $ & 5.658(2) & $4.05 \pm 0.07$  & 982  \\
$Co_{1.95}Ho_{0.05}MnSi     $ & 5.655(3) & $4.30 \pm 0.07$  & 983  \\
$Co_{1.90}Ho_{0.10}MnSi     $ & 5.659(2) & $3.94 \pm 0.07$  & 982  \\ \hline
\end{tabular}
\vspace{0.25cm}
\caption{Lattice parameters, saturation magnetizations and Curie temperatures 
for the studied compounds.}
\label{tab1}
\end{table}

\begin{table}[h]
\begin{tabular}{|c|c|c|c|c|c|c|c|}
\hline
\multirow{2}{*}{Sample} & Magnetic & \multicolumn{5}{c|}{Computed magnetic moments $(\mu_B)$}& Experimental$^*$ \\  \cline{3-7}
                        & coupling & M$_{Co}$ & M$_{Mn}$ & M$_{Ho}$ & M$_{Si}$ &   M$_{tot}$ & M(${\mu_B}/f.u.$) \\ \hline \hline
Co$_2$MnSi              &          & 1.003    &  3.026   &    -     &  -0.03   &    5.002    &   - \\
Co$_2$Ho$_{1/8}$Mn$_{7/8}$Si & P   & 1.013    &  2.866   &  9.779   &  -0.018  &    5.738    &   -  \\
Co$_2$Ho$_{1/8}$Mn$_{7/8}$Si & AP  & 1.009    &  2.866   & -9.979   &  -0.017  &    3.261    & 3.87 $\pm$ 0.2 \\
Co$_{15/8}$Ho$_{1/8}$MnSi & AP(Co1/Co2)& 0.941&  3.110   & -9.625   &  -0.047  &    3.626    & 3.76 $\pm$ 0.1 \\ \hline 
\end{tabular}
\vspace{0.25cm}
\caption{Calculated magnetic moments 
for different Ho-substituted compounds. 
The experimental data $^*$ are obtained 
by extrapolating the measured values
($x=0.05$ and $0.10$) to $x=0.125$ composition.}
\label{tab2}
\end{table}

\newpage 

\newpage 

\newpage

\end{document}